# SpeechJammer: A System Utilizing Artificial Speech Disturbance with Delayed Auditory Feedback

**Kazutaka Kurihara**

National Institute of Advanced Industrial Science and Technology

Chuo Dai 2, 1-1-1 Umezono,

Tsukuba, Ibaraki, Japan

k-kurihara@aist.go.jp

**Koji Tsukada**

Ochanomizu University / JST PRESTO

2-1-1 Otsuka Bunkyo-ku

Tokyo, Japan

tsuka@acm.org

## Abstract

In this paper we report on a system, "SpeechJammer", which can be used to disturb people's speech. In general, human speech is jammed by giving back to the speakers their own utterances at a delay of a few hundred milliseconds. This effect can disturb people without any physical discomfort, and disappears immediately by stop speaking. Furthermore, this effect does not involve anyone but the speaker. We utilize this phenomenon and implemented two prototype versions by combining a direction-sensitive microphone and a direction-sensitive speaker, enabling the speech of a specific person to be disturbed. We discuss practical application scenarios of the system, such as facilitating and controlling discussions. Finally, we argue what system parameters should be examined in detail in future formal studies based on the lessons learned from our preliminary study.

## ACM Classification Keywords
H.5.m. [Information interfaces and presentation (e.g., HCI)]: Miscellaneous

**General Terms**
Human Factors

**Introduction**
Speech and writing are two of the fundamental methods of communication between people. Of the two methods, communication based on speech has been widely used in daily life, even after the invention of writing, because it allows for communication using only the human body, and one-to-many broadcasting is easily achieved.

We live in the twenty first century, when it is said that communication is the most important means of resolving conflicts. However, there are still many cases in which the negative aspects of speech become a barrier to the peaceful resolution of conflicts, sometimes further harming society.

Here, we focus on two major categories of such negative features of speech; unavoidability and occupancy, which are defined as follows:

- Unavoidability: Speech can be initiated and continued by the speaker alone, and listener cannot avoid it.
- Occupancy: Speech can usually accept only one speaker at a time. If more than two people speak simultaneously, none of them are understandable (known as a "cross-talk" state).

We will discuss the following two typical situations in which the above features cause some problems.

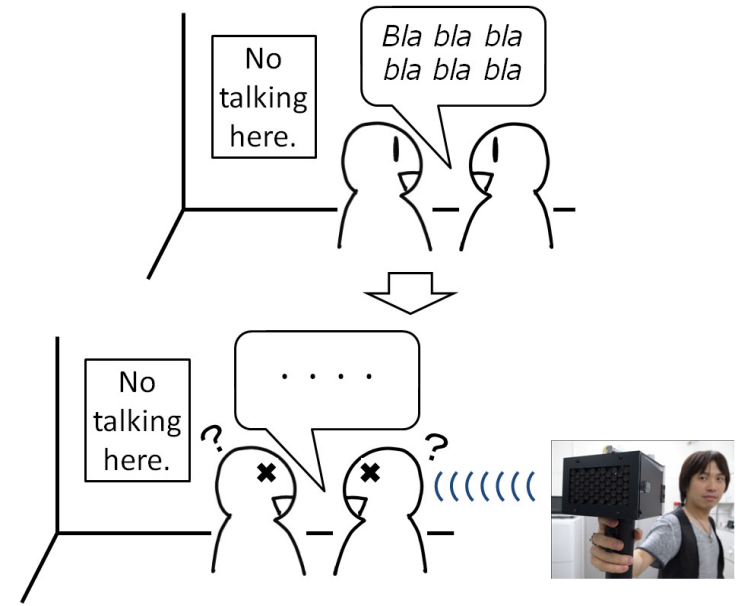

**figure 1.** Overview of SpeechJammer

*(1)"The louder, the stronger"*
Fair discussions are essential for resolving conflicts through communication. We have to establish and obey rules for proper turn-taking when speaking. However, some people tend to lengthen their turns or deliberately interrupt other people when it is their turn in order to establish their presence rather than achieve more fruitful discussions. Furthermore, some people tend to jeer at speakers to invalidate their speech. These are typical abuses of unavoidability and occupancy that allow "the louder, the stronger" to gain control of the proceedings.

*(2) "I cannot say `Be quiet!'"*
There are some public places where excessive talk is inappropriate, such as in libraries and trains [1].

We can infer that because speech has unavoidability, or cannot be easily avoided by listeners, we have established a consensus that we should not generate excessive levels of noise in public. However it is difficult to break off such inappropriate talk once it starts as we have to participate in the "inappropriate" conversation to tell the initial speakers that they should not do it. This contradiction may create a psychological burden. Even if we can cope with the burden, we may not achieve the initial goal of breaking off the talk without establishing peaceful turn-taking speech. Additional abuses of unavoidability and occupancy, discussed in (1), may occur and could lead to further conflict.

One typical passive solution for this situation is to use a headphone from a music player to invalidate the unavoidability and try to ignore the inappropriate behavior.

*Solution*
To cope with abovementioned typical situations, we focus on technologies that can control the properties of remote people's speech. As the first step in this aim, in this paper we report a system that jams remote people's speech using Delayed Auditory Feedback, a well-studied method involving the human auditory system (Figure 1).

This effect can disturb people without any physical discomfort, and disappears immediately the speaking stops. Furthermore, this effect does not involve anyone but the speaker. It is expected that the negative aspects of speech, which lead to all the problems mentioned above, can be relaxed by the ability to jam remote people's speech. Namely, unavoidability is relaxed, and we can control occupancy with proper turn-taking using the system.

In this paper, we first present the related work. After that, we introduce our proposed system and its implementation in detail. Finally, we argue the kind of system parameters that should be examined in detail in future formal studies based on lessons learned from our preliminary study.

**Related Work**
*Delayed Audio Feedback*
It is thought that when we make utterances we not only generate sound as output, but also we utilize the sound actually heard by our ears (called "auditory feedback") in our brains [5]. Our natural utterances are jammed when the auditory feedback is artificially delayed. It is thought that this delay affects some cognitive processes in our brain. This phenomenon is known as speech disturbance by Delayed Auditory Feedback (DAF).

DAF has a close relationship with stuttering. DAF leads physically unimpaired people to stutter; i.e., speech jamming. On the other hand, it is known that DAF can improve stuttering [1], and medical DAF devices are available [6]. We utilized DAF to develop a device that can jam remote physically unimpaired people's speech whether they want it or not. This device possesses one characteristic that is different from the usual medical

---
[1] We admit some cultural differences between countries.

DAF device; namely, the microphone and speaker are located distant from the target.

*Supporting Discussion*
This research focuses on controlling speech-based communications using a device. Previously, Nagao et al. studied the recording and reusing of discussions [4], and Sumi et al. tried to discover meaningful communication patterns via a bottom-up approach by the recording of many discussions [7]. There are also many studies on specific aspects of communication, such as decision-making by applying rules/constraints on communication, or facilitating communication through visualization of the properties of the discussion [2][3].

In this paper we develop a system that can impose a new strong constraint on speech-based discussion. Simply put, "makes speech difficult for some people." This constraint is thought to bring meaningful changes to communication patterns in discussions, and it also points the way to promising future research relating to discussion dynamics.

## The SpeechJammer System

*System Design*
To design "SpeechJammer," a system that jams people's speech, we first argue possible application scenarios of the system. The main design decision of the hardware is where its microphone and speaker should be fitted in the environment. The effect of DAF depends on the time required for transmission of the acoustic waves via the air. Therefore, we should carefully choose which parts of the transmission are electric, and which parts are transmitted via the air.

We suppose the application scenarios of the system as follow:

- Turn-taking controller for discussions in a meeting room
- Portable speech-jamming gun

These two scenarios correspond to the two typically problematic situations mentioned in the *Introduction* section. Figure 2 and 3 illustrate the hardware configurations for each scenario. In the figures, solid lines indicate the electric transmission of signals, wavy lines indicate the acoustic transmission of signals via the air, and blocks with *D* indicate the artificial delay components.

In the "turn-taking controller for discussions in a meeting room" scenario, we can utilize microphones equipped near every participant, and a number of public speakers in the room as the infrastructure (Figure 2). In this case, the signals are transmitted via the air in a one-way manner; i.e., only from the speakers to the participants.

In the "portable speech-jamming gun" scenario, the user disturbs inappropriate speech in public places using a gun-shape device equipped with a microphone and a speaker (Figure 3). In this case, the signals are transmitted via the air in a round-trip manner; i.e., between the gun user and the target.

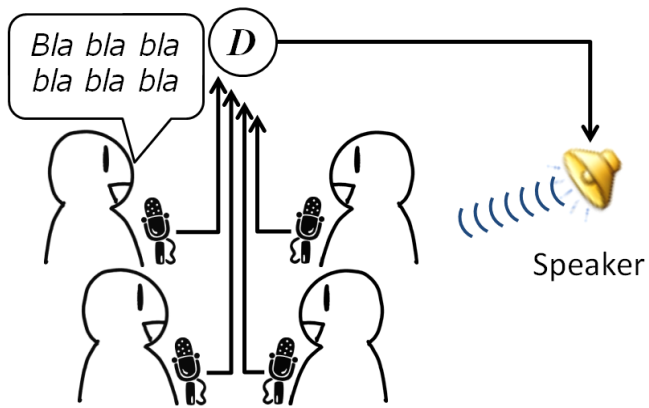

**figure 2.** Turn-taking controller for discussions in a meeting room.

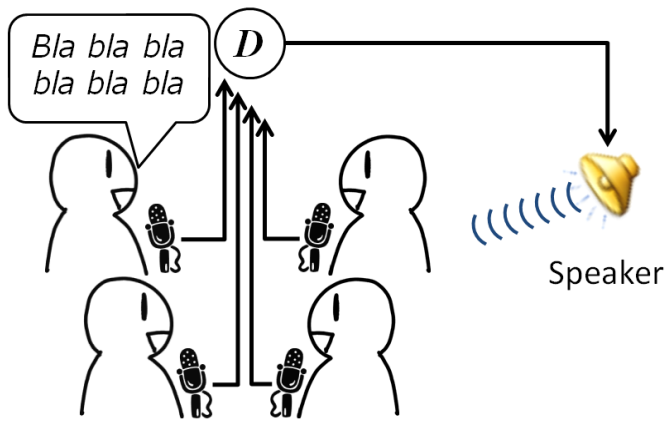

**figure 3.** Potable speech-jamming gun about supporting discussions.

In this paper, we propose a "portable speech-jamming gun" prototype SpeechJammer as this device has sufficient functional generality to be applied to the evaluation of the other scenario; i.e., the turn-taking controller for discussions in a meeting room, by directing guns toward every participant.

*Calculation of the Delay*
The SpeechJammer system has to deliver the speech back to the speaker with an appropriate, taking into consideration the distance between the speaker and the device.

Here we denote the distance between the speaker and the device as $x$[m], the air temperature as $t$[degrees Celsius], and the delay required for speech jamming using DAF as $D_{daf}$[sec], as shown in Figure 4. We can calculate the artificial delay $D$[sec] produced by the system by considering the speed of acoustic waves in 1[atm] air as follows:

$$D = D_{daf} - 2x/v \qquad (1)$$

Note that the speed of acoustic waves $v$[m/sec] is:

$$v = 331.5 + 0.61t \qquad (2)$$

From these equations we can say that if, for instance, we fix the delay $D_{daf}$ effectively for artificial stuttering as $D_{daf} = 0.2$[sec] in 20[degrees Celsius] air, $x \leq 34.37$[m] is obtained under the condition $D \geq 0$. This means that in 20[degrees Celsius] air the maximum available distance achievable by the system is about 34[m].

The SpeechJammer system can disturb the speech of any person within this maximum distance. When the application scenario allows the distance $x$ to be fixed, we can calculate a fixed $D$ in advance. When $x$ is unknown or varies, a distance sensor can measure $x$ and $D$ can then be calculated.

On the other hand, the SpeechJammer can disturb speech by a simpler setting with a fixed $D$ as follows. If we can assume that artificial stuttering occurs when $0.1 \leq D_{daf}$, by substituting $D=0.1$[sec] into Equation (1), we obtain $0 \leq x$, which means that a fixed $D$ is enough to jam speech at any distance. Note that as the relationship between $D_{daf}$ and its effect on the level of artificial stuttering, and the possible upper-bound of $D_{daf}$ are not clearly presented in the literature, we need to find these parameters through detailed experiments. Also note that, in the real world, the upper bound of $x$; i.e., the maximum available distance, is limited by the specifications of the microphone and speaker.

## Prototype Implementations

We implemented two SpeechJammer prototypes. Prototype #1 is the first version of SpeechJammer and has the basic functionalities (Figure 5). Figure 6 illustrates the system configuration.

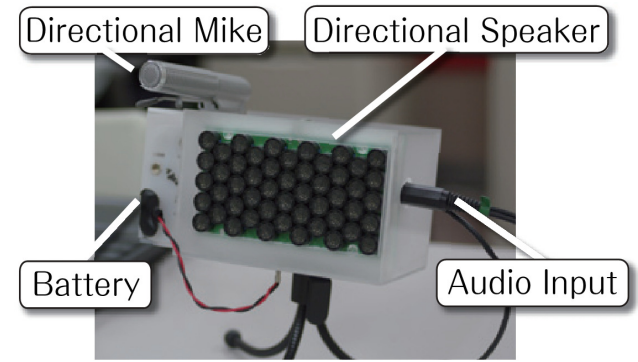

**figure 5.** Overview of SpeechJammer prototype #1.

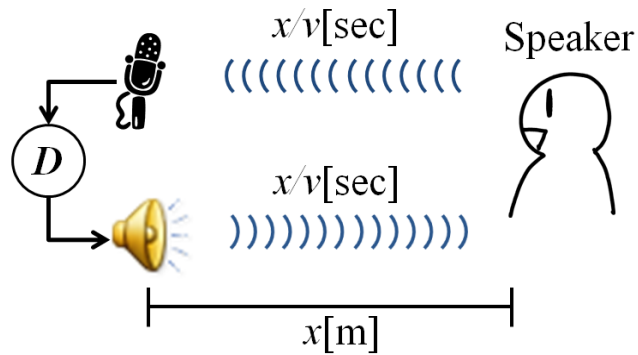

**figure 4.** Calculation of the delay $D$.

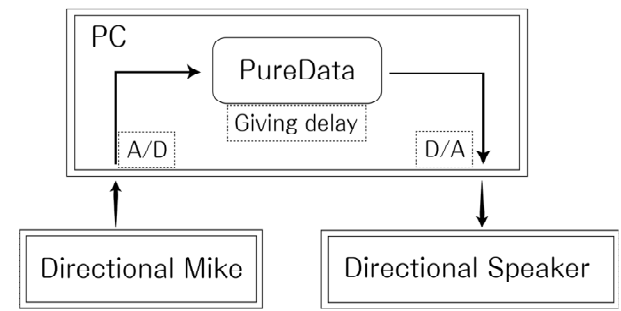

**figure 6.** System configuration of SpeechJammer prototype #1.

A direction sensitive microphone (Sony ECM-Z60) and a direction sensitive speaker (Tri-state Parametric Speaker Kit) are equipped on an acrylic case. The case has a screw thread for tripods and it can be attached on general tripods to adjust the angle, height, and so forth. The output terminal of the direction sensitive microphone is connected to the microphone input terminal of the host PC.

Similarly, the external input terminal is connected to the headphone output terminal of the host PC. The host PC runs an acoustic processing program written in Pure Data that gives arbitrary delays to the input from the direction sensitive microphone and outputs it to the direction sensitive speaker. Note that the device uses eight AA batteries as the power source to protect against noise interference from AC lines.

We checked the feasibility and the basic functionalities of SpeechJammer using the abovementioned prototype #1.

Next, we developed prototype #2, which is portable and enables standalone operations without a host PC (Figure 7, 8, and 9). Figure 10 illustrates the system configuration.

Besides a direction sensitive microphone (Sony ECM-CZ10) and a direction sensitive speaker (Tri-state Parametric Speaker Kit), a laser pointer, a distance meter, switches, and a mother board are fitted in an originally designed acrylic case.

The laser pointer is used to set the SpeechJammer's sight approximately, and is enabled and disabled by a switch on the back of the device.

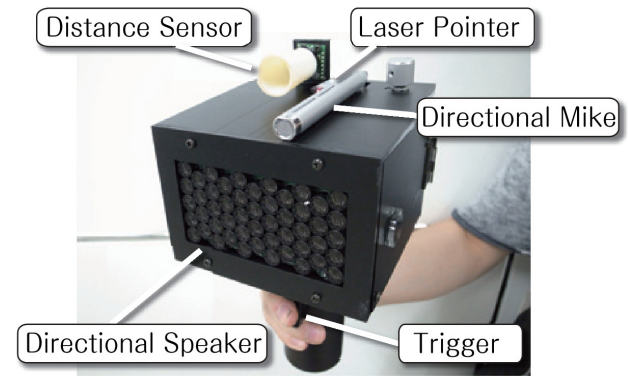

**figure 7.** Front view of SpeechJammer prototype #2.

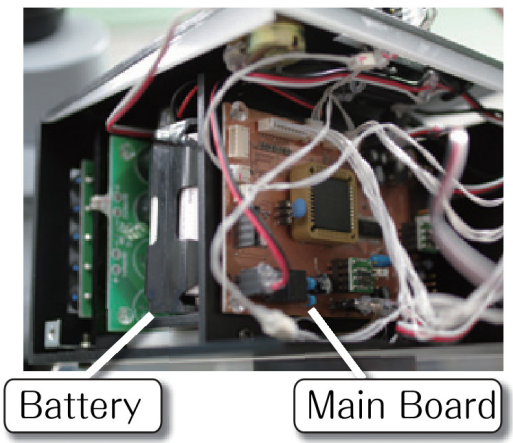

**figure 8.** Inside view of SpeechJammer prototype #2.

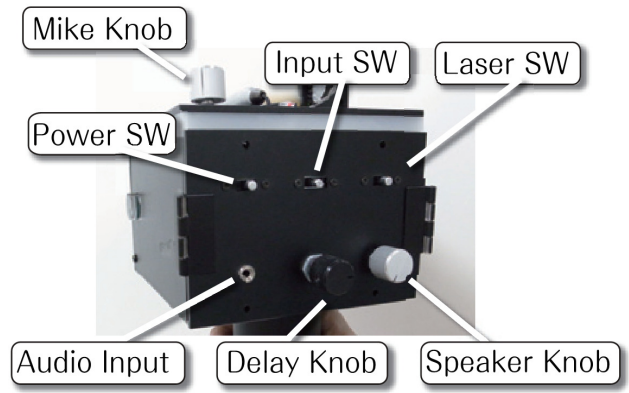

**figure 9.** Back views of SpeechJammer prototype #2.

The distance meter is used to measure distances between the target and the device for calculating the necessary artificial delays. A microcomputer (Microchip PIC18F-452), a digital delay IC (Rohm BU9262AFS), a pre-amplifier, a main amplifier, and auxiliary circuits are fitted on the mother board. The audio signals coming from the direction sensitive microphone are directed to the digital delay IC on the mother board via the pre-amplifier. The digital delay IC can be controlled using a serial interface and an appropriate delay from 9.2[msec] to 192[msec] is set. In this device, the delay is set using an 8-state rotary switch located on the back. We also prepared various modes using this rotary switch for future uses such as the automatic adjustment mode of the necessary delay using the distance meter, and modes enabling the periodic change of delays with various waveforms.

The output signals from the digital delay IC are directed to the direction sensitive speaker via the main amplifier. The pre-amplifier and the main amplifier are muted by default, and turned on when a trigger switch is pulled. The gain of the two amplifiers (i.e., the input gain and the output gain) can by adjusted using volume knobs located on the top and back of the case, respectively. The microcomputer is used for controlling the digital delay PC, the trigger switch, the rotary switch, the distance meter and so forth.

Consequently, users can easily operate the speech-jamming function by simply sighting the device toward the target and pulling the trigger switch like a pistol.

## Evaluation and Discussion

We conducted a preliminary study with five participants to examine the relationships between the various parameters of SpeechJammer and their effects on the artificial stuttering of participants. Prototype #1 was used in this study and the parameters were controlled on the host PC. Based on the lessons learned from the study, in the following sections we present the parameters that should be examined in detail in future formal studies.

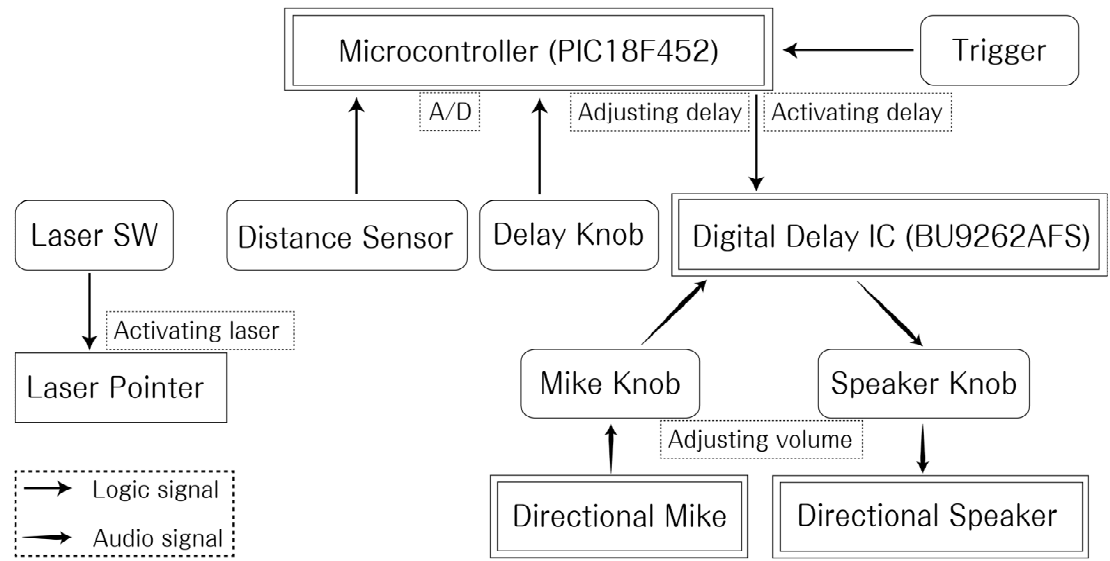

**Figure 10.** System configuration of SpeechJammer prototype #2.

*Delay Time*

Past studies relating to the improvement of stuttering using DAF at various delay *D* settings (between 0.004[sec] and 0.195[sec]) have been previously summarized [5]. From the summary it can be observed that stuttering improves relative to the size of the *D*. In our preliminary study, we obtained results consistent with those summarized (i.e., larger *D* values jam speech more effectively). We also observed speech jams caused by *D* values of more than 1[sec]. Considering these facts, it is necessary to examine the degrees of speech jamming at a wide range of the delay *D* values to obtain the optimum value.

*Time Varying Delay*

In our preliminary study, we examined not only fixed *D* values but also time variable *D* with sinusoids: $D=0.15+0.05\sin(2\pi T)$, denoting *T* as a time[sec] starting from zero.

From the results, we observed cases in which time variable *D* values have larger speech jamming effects than do fixed *D* values. Considering this, it is necessary to examine the degree of speech jamming at a wide variety of delay *D* values with different time series functions to obtain the optimum setting.

*Acoustic Gains*

In our preliminary study, the degree of speech jamming depended on the volume of the participants' voice, and the gain of the input/output amplifiers for the microphone and speaker, respectively. We can infer that the degree of speech jamming somehow relates to the gain ratio of the natural audio feedback given via the air/bone conductions and the artificial delayed audio feedback given by the system. Considering this, it is necessary to examine the degree of speech jamming at a wide range of acoustic gains, given by the amplifier volumes and the distance $x$, to obtain the optimum setting.

*Context of Speech*

Again, in our preliminary study, we dealt with "reading news aloud" and "spontaneous monologue" as the speech contexts. From the results, we observed a tendency for speech jamming to occur more frequently in the "reading news aloud" context than in the "spontaneous monologue" context. Further, it is obvious that speech jamming never occurs when meaningless sound sequences such as "Ahhh" are uttered over a long time period. Considering these facts, it is necessary to examine the degree of speech jamming in different speech contexts.

## Summary

In this paper we discussed two negative features of speech, unavoidability and occupancy, which can be barriers toward peaceful communication. We then developed two prototypes of the SpeechJammer system with the aim of relaxing such negative features using DAF. The system can disturb remote people's speech without any physical discomfort. In the future, we will conduct detailed evaluations of the system to clarify the relationships between various parameters of the system and its effect on the level of artificial stuttering.

## Acknowledgments

This research project is supported by JST PRESTO.